# EFFECT OF CHEMICAL PRESSURE ON THE MAGNETIC TRANSITION OF MULTIFERROIC $Bi_{1-x}Ca_xFeO_3$

G. Catalan[1], K. Sardar[2], N. S. Church[1], J. F. Scott[1], R. J. Harrison[1], S. A. T. Redfern[1]

[1]Department of Earth Sciences, University of Cambridge, Downing Street, Cambridge CB2 3EQ, United Kingdom.

[2]Department of Chemistry, University of Warwick, Coventry CV4 7AL, United Kingdom.

**Abstract**

Multiferroic $BiFeO_3$ ceramics have been doped with Ca. The smaller ionic size of Ca compared with Bi means that doping acts as a proxy for hydrostatic pressure, at a rate of 1%Ca=0.3GPa. It is also found that the magnetic Neel temperature ($T_{Neel}$) increases as Ca concentration increases, at a rate of 0.66K per 1%Ca (molar). Based on the effect of chemical pressure on $T_{Neel}$, we argue that applying hydrostatic pressure to pure $BiFeO_3$ can be expected to increase its magnetic transition temperature at a rate around $\partial T_N / \partial P \sim 2.2$K/GPa. The results also suggest that pressure (chemical or hydrostatic) could be used to bring the ferroelectric critical temperature, Tc, and the magnetic $T_{Neel}$ closer together, thereby enhancing magnetoelectric coupling, provided that electrical conductivity can be kept sufficiently low.

## Introduction

Bismuth ferrite $BiFeO_3$ (BFO) is arguably the best studied- magnetoelectric multiferroic oxide at the moment[1], the interest being due to the fact that both the magnetic and ferroelectric ordering take place well above room temperature, with the ferroelectric polarization being the largest of any ceramic (almost $100\mu C/cm^2$ along the polar <111> direction [2, 3]). Because of this, it has received an enormous amount of attention, and new features of its behaviour and phase transitions are constantly being uncovered. The temperature-pressure phase diagram of $BiFeO_3$, in particular, has proved to be more complex than initially thought, with several new phase transitions being reported just in the past year [1, 4, 5, 6, 7].

One problem to study the phase diagram, though, is that key phase transitions such as the ferroelectric-paraelectric one and the metal-insulator one take place at either very high temperatures ($T_C$ =1100K and $T_{MI}$ =1200K respectively) or pressures (10GPa and 50 GPa respectively), the latter being quite hard to achieve in most labs. Because of this, it is useful to extend studies to doped specimens, where the different size of the dopant could have an analogous effect to that of pressure (this is sometimes called "chemical pressure"). Here we have doped $BiFeO_3$ with Ca, which has a smaller ionic size than Bi and may therefore be expected to act as a good proxy for hydrostatic pressure. Ca-doped BFO is indeed beginning to attract attention [8, 9, 10], but so far there has been no link between the properties of this compound and those of of pure BFO under pressure, nor is it known how doping affects the magnetic properties. We have found that Ca doping increases the magnetic transition temperature ($T_{Neel}$) while decreasing the volume of the unit cell, which suggests that hydrostatic pressure should also lead to an increase of $T_{Neel}$ in pure $BiFeO_3$.

## Experiment

The ceramic samples of Ca-BFO were synthesised following the recipe proposed by Ghosh *et al*[11]. 1:1 molar ratio $Bi(NO_3)_3$ and $Fe(NO_3)_3.9H_2O$ was dissolved in water. To it, tartaric acid (molar ratio of metal to tartaric acid = 1:1) was added to obtain a clear yellow-coloured solution. was added to obtain a clear yellow colour solution. The solution was evaporated at 100 °C (boiling condition) under constant stirring, in order to obtain a brownish solid precursor. The solid precursor was ground to make powder which was first dried in air in an oven at 100 °C for another 24 hours. Calcium doping was achieved by adding stoichiometric amount of calcium nitrate in the starting solution. The precursor powder was then calcined in air at 600 °C for 2 h. Pellets prepared from the calcined powder was sintered at 700 °C for 3 h in air. It was observed that with increasing amount of calcium, the sintering temperature could be extended up to 850 °C without any noticeable phase separation. However, the sintering temperatures of all the samples were kept the same for the sake of comparison of physical properties.

The lattice parameters of the ceramic were determined by x-ray powder diffraction at room temperature using a Bruker D8 diffractometer with Cu-K$_{\alpha 1}$ radiation. Diffraction patterns were collected in $\theta$-$2\theta$ geometry from °$2\theta$=10 to 150°. Lattice parameters were obtained from Rietveld refinement of the measured diffraction patterns using GSAS. The magnetic ordering temperature was determined from specific heat and low-field AC susceptibility measurements. Heat flow measurements were made through the Neel transition using a PE Diamond DSC. Around 10mg of powdered sample of each composition was enclosed in a Al sample can and run between room temperature and 773 K at a controlled heating rate with dry nitrogen purge. Temperature was calibrated against the melting temperatures of indium and zinc standards. Data from ten runs were combined to reduce noise, and poor quality scans were eliminated from the data set. Neel transition is accompanied by a significant peak in heat flow on heating, and the transition temperature was obtained from the point at which the first derivative of heat flow passed through zero. Magnetic susceptibility measurements were carried out using an AGICO MKF1-FA kappabridge with an AC field of 200 A/m. Specimens were measured 10 times up to ~440 °C (713K) in air with a heating rate of 10K/min. The Neel temperature was measured as the peak in the susceptibility signal on the heating run, as methods which determine the transition from changes in gradient were imprecise due to noise.

**Results**

The structure of BFO is rhombohedral at room temperature, and the influence of Ca-doping is to reduce both the volume and the rhombohedral distortion of the unit cell. Figure 1 shows the pseudocubic lattice parameter of rhombohedral Ca-BFO, plotted as a function of Ca concentration. As expected, the pseudocubic lattice parameter (defined as the cube root of the perovskite unit cell volume) decreases with increasing Ca content. From a linear fit of the data, we obtain that the lattice parameter decreases at a rate of -0.003Å per 1% mol of Ca doping. We have analysed the BFO pressure dependence of unit cell volume reported by Gavriliuk *et al.* [5] and found that the unit cell lattice parameter decreases at a rate of ca. -0.01Å/GPa (we have only fitted pressures below ~10GPa to avoid the high pressure phase transitions [5, 7]). Based on this result, we make the association that, structurally at least, 1%Ca = 0.3GPa of "chemical pressure".

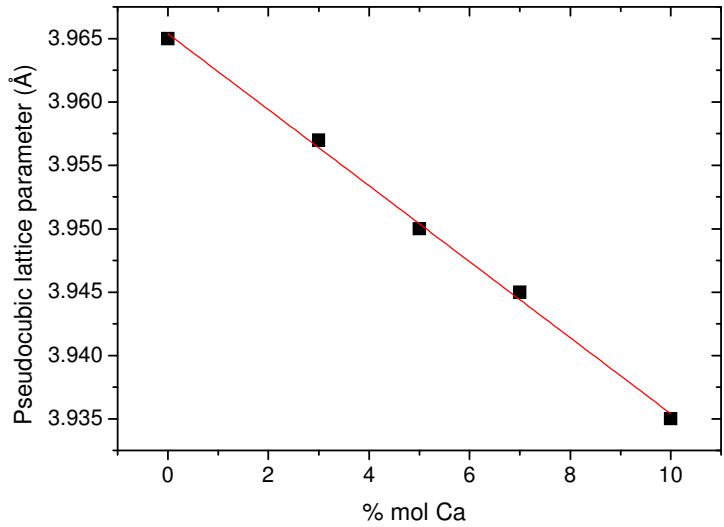

**Figure 1.** Evolution of the perovskite pseudocubic unit cell as a function of Ca concentration. The red line is a least squares fit yielding a compression rate of -0.003Å per 1%mol of Ca.

We now turn to the effect of doping on the Neel temperature. Figure 2 shows a typical differential scanning calorimetry (DSC) measurement for a Ca-doped sample (7% Ca in this case); the Neel temperature shows up as a distinct peak. The inset of the figure shows the Ca-concentration dependence of $T_{Neel}$ as determined from the specific heat measurements. It is found that $T_N$ increases at about 0.66 K per 1%Ca. Combining this result with the chemical pressure equivalence (1%Ca=0.3GPa) we find that $\partial T_N / \partial P$ ~2.2 K/GPa of chemical pressure.

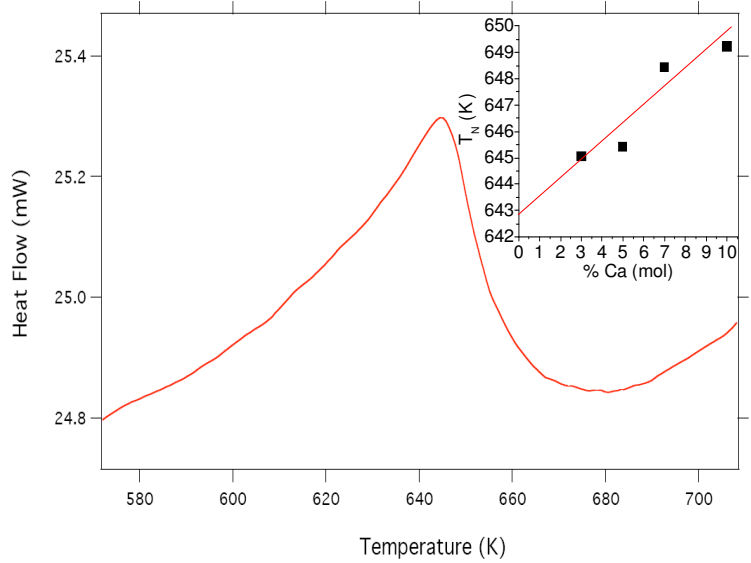

**Figure 2**. Differential scanning calorimetry of 7%Ca sample, showing the clear peak in specific heat at the Neel temperature. Inset: $T_{Neel}$ as determined from specific heat as a function of Ca concentration.

Figure 3 shows the magnetic susceptibility. Again, a distinct peak can be seen which we identify with $T_{Neel}$. Plotting this as a function of Ca concentration yields roughly the same result as the calorimetry measurements, although the 5% sample is off the trend in the magnetic measurements; this sample was observed to have an impurity phase of hematite ($Fe_2O_3$), so it is considered less reliable in terms of both stoichiometry and magnetic signal, given the strong magnetism of hematite. From a linear fit of the data, we obtain that the Neel temperature increases at a rate of 0.6K per 1% Ca, or $\partial T_N / \partial P \sim 2.0$ K/GPa of chemical pressure, which agrees well with the result obtained from calorimetry.

How do these results compare with the effect of real hydrostatic pressure? We know of no studies to date on the effect of hydrostatic pressure on the Neel temperature of $BiFeO_3$, but for perovskite orthoferrites the rate is [12] $\partial T_N / \partial P = 4 - 7$ K/GPa, which is comparable with, though somewhat bigger than, the effect of chemical pressure found in this work ($\partial T_N / \partial P \sim 2$K/GPa). We do not know at this stage whether the difference is due to the fact that chemical pressure is not identical to hydrostatic pressure (specifically, the non isovalent nature of the doping, $Ca^{2+}$ for $Bi^{3+}$, affects the electronic structure) or to the fact that the crystal structures of orthoferrites and BFO are different (orthoferrites are orthorhombic, BFO is rhombohedral). Direct measurements of hydrostatic pressure should answer that.

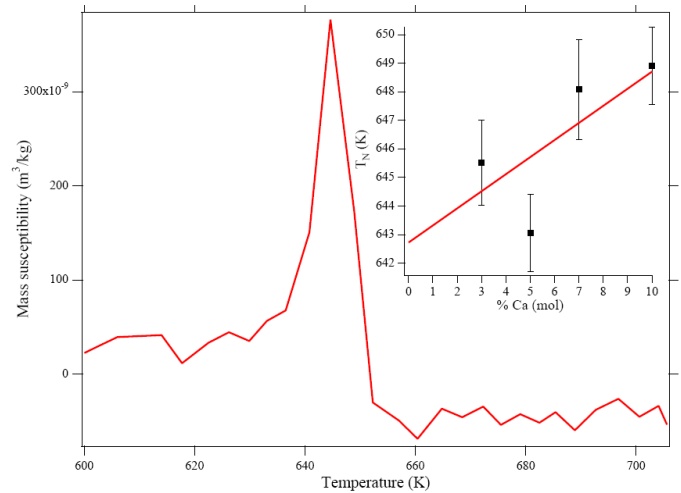

**Figure 3.** Magnetic AC susceptibility measurement for BFO-3%Ca, showing the distinct peak at the Neel temperature. Inset: linear fit of $T_N$ as a function of composition.

**Discussion**

$Ca^{2+}$ is, like $Bi^{3+}$, a non-magnetic ion, so it is not expected to directly contribute to the magnetic properties of this material. However, due to the different valence, doping BFO with Ca introduces a charge imbalance that must be equilibrated either by oxygen vacancies or by a change in the valence of iron from 3+ to 4+. Both charge compensation mechanisms exist in nature and may coexist in our samples. The brownmillerite structure ($CaFeO_{2.5}$) represents one extreme where all iron ions are $Fe^{+3}$ and charge neutrality has been preserved purely by oxygen vacancies[13, 14]. On the other hand, vacancy-free $CaFeO_3$ can be made[15, 16, 17] with all irons in the 4+ state or a charge disproportionation of $Fe^{3+}$ and $Fe^{5+}$. These two situations should in principle lead to different outcomes for the magnetic ordering temperature. $Fe^{+4}$ has one less valence electron than $Fe^{+3}$ and therefore the magnetic interactions should be weaker and the ordering temperature lower than for the $Fe^{+3}$ compound; conversely, the pure $Fe^{+3}$ compound, $CaFeO_{2.5}$, has a higher $T_{Neel}$ than $BiFeO_3$ (720K instead of 643K) [14].

Let us first consider the first possibility: that charge neutrality is preserved by oxygen vacancies so that $Fe^{+3}$ does not change its valence state and nor does the magnetic exchange constant. A simple mix rule, then, suggests that the Neel temperature should evolve as the weighted average between the transition temperatures of $BiFeO_3$ and $CaFeO_3$:

$$T_{Neel} = X T_{Neel}^{CaFeO_{2.5}} + (1-X) T_{Neel}^{BiFeO_3} \qquad (1)$$

Where X is the molar concentration of Ca in the mixture. According to this, the Neel temperature should increase 0.77K per 1% mol of Ca doping. This is comparable to, though somewhat bigger than, our experimentally measured values of 0.6-0.66 K per 1% mol Ca, suggesting that charge compensation by oxygen vacancies is quite plausible.

We now turn to charge compensation by tetravalent iron. Pure $CaFeO_3$ has a very low $T_{Neel}$= 125K [16], and one may therefore think that $Fe^{+4}$ is not likely to be present in our samples given that they have an increased $T_{Neel}$ instead. However, the picture is somewhat complicated by the fact that $CaFeO_3$ has concomitant charge-ordering and metal-insulator transitions at the magnetic transition[15, 16, 17], so it is not clear whether the "real" Neel temperature would have been bigger where it not frustrated by the electron delocalization of the metallic state (Pauli paramagnetism), a situation that has also been observed in perovskite nickelates [18]. Furthermore, the hypothetical change of valence in our samples from $Fe^{+3}$ to $Fe^{+4}$ can indirectly reinforce the magnetism through straightening the Fe-O-Fe bond angle, as explained below.

Though $BiFeO_3$ has a rather exotic inconmensurate magnetic structure [19], its local spin structure is a G-type antiferromagnet, meaning that, on a local level, the magnetic properties of BFO are comparable to those of perovskite orthoferrites [20, 21]. In these, the strength of the antiferromagnetic superexchange interaction depends the Fe-O-Fe angle ($\equiv \theta$); specifically, it is proportional to $\cos\theta$ [22, 23, 24]. In a perfectly cubic perovskite, $\theta$ would be exactly 180° and the antiferromagnetic coupling would be maximized. For pure BFO, however, this angle is $\theta \sim 155°$ [25, 26]. The buckling of the Fe-O-Fe bond angle is itself due to the size mismatch between the ions, which can be quantified using the tolerance factor [27]:

$$t = \frac{(r_{Bi} + r_O)}{\sqrt{2}(r_{Fe} + r_O)} \qquad (2)$$

This is 1 for perfectly cubic perovskites, but for BFO it is smaller ($t$~0.887). In the orthoferrites, A-site substitution can increase $t$, thereby straightening $\theta$ and increasing $T_{Neel}$ [22, 23, 24]. However, while isovalent substitution requires an increase in A-site ionic size in order to straighten $\theta$, non-isovalent substitution does not, because the charge imbalance may be compensated by a change from $Fe^{3+}$ to $Fe^{4+}$: since the latter has a smaller ionic size, the denominator in eq. (2) can decrease thus making $t$ increase towards 1 (consequently straightening $\theta$). Nevertheless, for the case of $Ca^{+2}$ substituting for $Bi^{+3}$, the increase in $t$ is rather small (~1% using the ionic sizes from Shannon [29]). Using the empirical relationship between tolerance factor and octahedral tilting of Megaw and Darlington [28], we can estimate that such a small change in $t$ would lead to only ~1° change in octahedral rotation, or ~2° increase in the exchange angle $\theta$. Given the relationship between $\theta$ and $T_{Neel}$ for orthoferrites [22, 23], an increase of $T_{Neel}$ by only 0.2K per 1% Ca can thus be expected. This is only one third of the increase rate measured in our samples, suggesting that tetravalent iron is unlikely to be the main charge compensation mechanism.

The effect of doping may also have useful consequences. As shown here, chemical pressure increases magnetic ordering temperature. But, at the same time, pressure decreases the ferroelectric $T_C$: the paraelectric $\beta$ phase above 1100K is the same as the orthorhombic phase above 10GPa at room temperature [4, 7, 30, 31]. For the right amount of pressure or doping, then, one may expect the two ferroic critical temperatures may coincide, leading to a maximization of magnetoelectric coupling. The Neel temperature of BFO is 643K, and it increases at a rate of 2.2K/GPa. The ferroelectric temperature, on the other hand, is 1100K at ambient pressure and it decreases, roughly, at a rate of 80K/GPa. The ferroelectric and magnetic ordering temperatures of pure $BiFeO_3$ should therefore coincide when 643+2.2P=1100-80P (with P expressed in GPa), i.e., when P~5.5GPa (or a doping concentration of 18%Ca). We note parenthetically that a pressure of 5.5GPa is in the the range in which an intermediate monoclinic phase has been reported [7], and one may speculate whether this has to do with the increased magnetoelectric coupling predicted here. The calculation for critical doping, however, assumes that the Ca doping does not itself significantly affect the chemical basis of the ferroelectricity, which is not true: ferroelectricity in BFO depends on the lone-pair polarization of the $Bi^{+3}$ ion, which is absent in $Ca^{2+}$, so 18% is probably an overestimate of the Ca concentration required and Tc may decrease more rapidly with doping than with pure pressure.

Unfortunately, a negative side effect of doping in our samples is that their conductivity becomes too high for the magnetoelectric properties to be directly measured. On the issue of conductivity, we note that $BiFeO_3$ has a metal insulator transition as a function of pressure and/or temperature [4, 5]. This MI transition is thought to be due to gradual closing of the charge transfer gap between O and Fe [1, 4, 31]. The charge transfer gap is itself directly correlated to the orbital overlap between the oxygen *p* states and the iron *d* states, the overlap being also bigger when the Fe-O-Fe angle is straighter [31]. Accordingly, it is expected that i) the bandgap of Ca-doped BFO should be considerably lower than that of pure BFO, leading to much increased conductivity and ii) the critical temperature of the metal insulator transition should also decrease with Ca doping. This is consistent with the observation that pure $CaFeO_3$ displays a metal-insulator transition at ambient pressure and $T_{MI}$~115K[16], compared with $T_{MI}$~1203K for pure $BiFeO_3$[4]. Using our calculated correlation between doping and pressure, one may regard $CaFeO_3$ as structurally similar to applying 33GPa to $BiFeO_3$, which is not far off the actual hydrostatic pressure required to induce the MI transition in BFO at low temperature [5, 6].

**Conclusions**

In summary, Ca-doping contracts the lattice of $BiFeO_3$ and isn in this respect equivalent to applying pressure. The correlation between chemical pressure and increase in Neel temperature is consistent with a straightening of the Fe-O-Fe exchange angle, an effect well known in perovskite orthoferrites [12, 22, 23, 24]. Based on this, it is argued here that chemical pressure can in principle be used as an effective means by which to tune the ferroelectric and magnetic transition temperatures so as to make them coincide, thereby enhancing magnetoelectric coupling. We expect the antiferromagnetic Neel temperature and the ferroelectric Curie temperature to coincide at pressures of the order of ~5.5 GPa, or a doping concentration of ~18% mol Ca. However, due to the correlation between charge transfer gap and Fe-O-Fe angle [31], we also expect that Ca-doping will reduce both the conduction bandgap and the metal-insulator transition temperature of BFO. Furthermore, the charge imbalance introduced by the non-isovalent doping can be compensated by oxygen vacancies, and these act as charge donors that further increase conductivity. Any doping strategy aimed at increasing the magnetoelectric coupling via structural tuning of the exchange angle will therefore have to deal with the problem increased conductivity first.

**References**

1. G. Catalan, J. F. Scott, Advanced Materials (in press 2008).

2. J. Wang, J. B. Neaton, H. Zheng, V. Nagarajan, S. B. Ogale, B. Liu, D. Viehland, V. Vaithyanathan, D. G. Schlom, U. V. Waghmare, N. A. Spaldin, K. M. Rabe, M. Wuttig, and R. Ramesh, Science **299**, 1719 (2003).

3. D. Lebeugle, D. Colson, A. Forget, M. Viret, P. Bonville, J. F. Marucco, S. Fusil, Appl. Phys. Lett. **91**, 022907 (2007).

4. R. Palai, R. S. Katiyar, H. Schmid, P. Tissot, S. J. Clark, J. Robertson, S. A. T. Redfern, G. Catalan, and J. F. Scott, Phys. Rev. B **77**, 014110 (2008).

5. G. Gavriliuk, V. V. Struzhkin, I. S. Lyubutin, S. G. Ovchinnikov, M. Y. Hu and P. Chow, Phys. Rev. B **77**, 155112 (2008).

6. O. E. González-Vázquez and J. Íñiguez, Phys. Rev. B **79**, 064102 (2009).

7. R. Haumont, P. Bouvier, A. Pashkin, K. Rabia, S. Frank, B. Dkhil, W. A. Crichton, C. A. Kuntscher, J. Kreisel, arXiv:0811.0047 (2008).

8. V. A. Khomchenko, D. A. Kiselev, J. M. Vieira, A. L. Kholkin, M. A. Sá and Y. G. Pogorelov., Appl. Phys. Lett. **90**, 242901 (2006).

9. D. Kothari, V. Raghavendra Reddy, A. Gupta, V. Sathe, and A. Banerjee, S. M. Gupta and A. M. Awasthi, Appl. Phys. Lett. **91**, 202505 (2007).

10. O. Troyanchuk, D. V. Karpinsky, M. V. Bushinskii, O. Prokhnenko, M. Kopcevicz, R. Szymczak, and J. Pietosaand, Journal of Experimental and Theoretical Physics **107**, 83 (2008).

11. S. Ghosh, S. Dasgupta, A. Sen, and H. S. Maiti, J. Amer. Ceram. Soc. **88** 1349 (2005).

12. N. A. Halasa, G. DePasquali, H.G. Drickamer, Phys. Rev. B **10**, 154 (1974).

13. E. F. Bertaut, P. Blum and A. Sagnieres, Acta Cryst. **12**, 149 (1959).

14. T. Takeda, Y. Yamaguchi, S. Tomiyoshi, M. Fukase, M. Sugimoto, and W. Watanabe, J. Phys. Soc. Jpn. **24**, 446 (1968).

15. F. Kanamaru, H. Miyamoto, Y. Mimura, M. Koizumi, M. Shimada, S. Kume and S. Shin, Mat. Res. Bull. **5**, 257 (1970).

16. Y. Takeda, S. Naka, M. Takano, T. Shinjo, T. Takada, M. Shimada, Mat. Res. Bull **13**, 61 (1978).

17. M. Takano, S. Nasu, T. Abe, K. Yamamoto, S. Endo, Y. Takeda, J.B. Goodenough Phys Rev Lett **67**, 3267 (1991).

18. G. Catalan, Phase Transitions **81**, 729 (2008).

19. I. Sosnovska,, T. Peterlin-Neumaier, E. Steichele, J. Phys. C **15**, 4835 (1982).

20. C. Ederer, N. Spaldin, Phys. Rev. B **71**, 60401 (2005).

21. M. K Singh, R. S Katiyar and J F Scott, J. Phys.: Condens. Matter **20**, 252203 (2008).

22. D. Treves, M. Eibschutz, and P. Coppens, Phys. Lett. **18**, 216 (1965).

23. A. Bombik, B. Lesniewska, J. Mayer, A. W. Pacyna, J. Magnetism and Magnetic Materials **257**, 206 (2003).

24. J. S. Zhou, J. B. Goodenough, Phys. Rev. B **77**, 132104 (2008).

25. F. Kubel and H. Schmid, Acta Crys. B **46**, 698 (1990).

26. Palewicz, R. Przeniosło, I.Sosnowska and A. W. Hewat, Acta Cryst. B **63**, 537 (2007).

27. V. M. Goldschmidt, Naturwissenschaften **14**, 477 (1926).

28. H. D. Megaw, C. N. W. Darlington, Acta Cryst. A **31**, 161 (1975).

29. R. D. Shannon, Acta Cryst. A **32**, 751 (1976).

30. D. C. Arnold, K. S. Knight, F. D. Morrison, P. Lightfoot, Phys. Rev. Lett. **102**, 027602 (2009).

31. S. A. T. Redfern, J. N. Walsh, S. M. Clark, G. Catalan, J. F. Scott, arXiv:0901.3748v2. (2009).